\documentclass[a4paper,clock]{article}
\usepackage[dvips]{epsfig}
\usepackage{amssymb}
\usepackage{bm}
\usepackage{dcolumn}
\usepackage{amsmath}
\usepackage{slashed}

\catcode`\@=11
\def\marginnote#1{}
\newcount\hour
\newcount\minute
\newtoks\amorpm
\hour=\time\divide\hour by60
\minute=\time{\multiply\hour by60 \global\advance\minute
by-\hour}\edef\standardtime{{\ifnum\hour<12
\global\amorpm={am}%
        \else\global\amorpm={pm}\advance\hour by-12 \fi
        \ifnum\hour=0 \hour=12 \fi
        \number\hour:\ifnum\minute<10
0\fi\number\minute\the\amorpm}}
\edef\militarytime{\number\hour:\ifnum\minute<10
0\fi\number\minute}

\def\draftlabel#1{{\@bsphack\if@filesw {\let\thepage\relax
   \xdef\@gtempa{\write\@auxout{\string
      \newlabel{#1}{{\@currentlabel}{\thepage}}}}}\@gtempa
   \if@nobreak \ifvmode\nobreak\fi\fi\fi\@esphack}
        \gdef\@eqnlabel{#1}}
\def\@eqnlabel{}
\def\@vacuum{}
\def\draftmarginnote#1{\marginpar{\raggedright\scriptsize\tt#1}}
\def\draft{\oddsidemargin -.5truein
        \def\@oddfoot{\sl preliminary draft \hfil
        \rm\thepage\hfil\sl\today\quad\militarytime}
        \let\@evenfoot\@oddfoot \overfullrule 3pt
        \let\label=\draftlabel
        \let\marginnote=\draftmarginnote

\def\@eqnnum{(\theequation)\rlap{\kern\marginparsep\tt\@eqnlabel}%
\global\let\@eqnlabel\@vacuum}  }

\def\numberbysection{\@addtoreset{equation}{section}
        \def\theequation{\thesection.\arabic{equation}}}

\def\underline#1{\relax\ifmmode\@@underline#1\else
 $\@@underline{\hbox{#1}}$\relax\fi}

\catcode`@=12
\relax

\numberbysection

\advance \topmargin by -\headheight
\advance \topmargin by -\headsep

\evensidemargin \oddsidemargin
\def\br{\begin{eqnarray}}
\def\er{\end{eqnarray}}
\def\be{\begin{equation}}
\def\ee{\end{equation}}

\def\({\left(}
\def\){\right)}

\relax

\def\a{\alpha}

\def\b{\beta}

\def\d{\delta}

\def\g{\gamma}

\def\pa{\partial}

\def\s{\sigma}

\def\tp0{\Theta_{+}^{(0)}}
\def\tm0{\Theta_{-}^{(0)}}

\def\vp{\varphi}

\def\f#1#2#3 {f^{#1#2}_{#3}}

\def\win1{{\sf w_{1+\infty}}}

\def\Win1{{\sf W_{1+\infty}}}

\def\rlx{\relax\leavevmode}
\def\inbar{\vrule height1.5ex width.4pt depth0pt}
\def\IZ{\rlx\hbox{\sf Z\kern-.4em Z}}
\def\IR{\rlx\hbox{\rm I\kern-.18em R}}
\def\IC{\rlx\hbox{\,$\inbar\kern-.3em{\rm C}$}}
\def\IN{\rlx\hbox{\rm I\kern-.18em N}}
\def\IO{\rlx\hbox{\,$\inbar\kern-.3em{\rm O}$}}
\def\IP{\rlx\hbox{\rm I\kern-.18em P}}
\def\IQ{\rlx\hbox{\,$\inbar\kern-.3em{\rm Q}$}}
\def\IF{\rlx\hbox{\rm I\kern-.18em F}}
\def\IG{\rlx\hbox{\,$\inbar\kern-.3em{\rm G}$}}
\def\IH{\rlx\hbox{\rm I\kern-.18em H}}
\def\II{\rlx\hbox{\rm I\kern-.18em I}}
\def\IK{\rlx\hbox{\rm I\kern-.18em K}}
\def\IL{\rlx\hbox{\rm I\kern-.18em L}}
\def\one{\hbox{{1}\kern-.25em\hbox{l}}}
\def\0#1{\relax\ifmmode\mathaccent"7017{#1}%
B        \else\accent23#1\relax\fi}

\def\PRD#1#2#3{{\sl Phys. Rev.} {\bf D#1} (#2) #3}

\def\JPAMT#1#2#3{{\sl J. Physics A: Math. Theor.} {\bf A#1} (#2) #3}

\def\JHEP#1#2#3{{\sl JHEP} {\bf #1} (#2) #3}

\def\JPAMG#1#2#3{{\sl J. Physics A: Math. Gen.} {\bf A#1} (#2) #3}

\def\EPJC#1#2#3{{\sl Eur. Phys. J. C} {\bf #1} (#2) #3}
 
\def\MoC#1#2#3{{\sl MATHEMATICS OF COMPUTATION} {\bf #1} (#2) #3}
\def\ChPB#1#2#3{{\sl Chinese Phys. B} {\bf #1} (#2) #3}
 
\begin{document}
\begin{titlepage}

\vskip .6in

\begin{center}
{\large {\bf Heun-function analysis of the Dirac spinor spectrum in a sine–Gordon soliton background}}
\end{center}

\normalsize
\vskip .4in

\begin{center}

H. Blas, R.P.N. Laeber Fleitas and  J. Silva Barroso 

\par \vskip .2in \noindent

$^{a}$ Instituto de F\'{\i}sica\\
Universidade Federal de Mato Grosso\\
Av. Fernando Correa, $N^{0}$ \, 2367\\
Bairro Boa Esperan\c ca, Cep 78060-900, Cuiab\'a - MT - Brazil. \\ 
\normalsize
\end{center}
\par \vskip .3in \noindent
 
We study the Dirac spectrum in a sine-Gordon soliton background, where the induced position-dependent mass reduces the spectral problem to a Heun-type differential equation. Bound and scattering sectors are treated within a unified framework, with spectral data encoded in Wronskians matching local Heun solutions and exhibiting explicit dependence on the soliton parameters and the bare fermion mass. This formulation enables a systematic analysis of spinor bound and scattering states, supported by analytic and numerical verification of wave function matching across the soliton domain. The present work is related to arXiv:2512.07658 and emphasizes a pedagogical treatment of scattering states within the Heun-equation formalism.  

\end{titlepage}

\section{Introduction}

Kink-fermion systems continue to attract significant interest due to their importance in nonperturbative quantum field theory \cite{raja}. These systems exhibit fermionic zero modes, charge fractionalization, and excited bound states, and are studied either via numerical approaches that include fermion back-reaction or within frameworks assuming a fixed kink background; for exact and self-consistent treatments of fermionic back-reaction, see \cite{prd2, arxiv} and references therein.

Topological soliton–fermion systems lead naturally to spectral problems governed by second-order differential equations with multiple singularities. In the presence of a sine–Gordon soliton background, the Dirac equation reduces to a Fuchsian system that can be systematically mapped into the canonical Heun equation. This structure reflects the underlying multi-scale and topological features of the interaction and provides a unified analytic setting for bound and scattering states.

On the other hand, Heun's equation has become increasingly pivotal in modern theoretical and mathematical physics, offering a unifying framework for the analysis of a diverse range of complex phenomena that extend well beyond the capabilities of simpler special functions, such as the hypergeometric functions. It is the most general form of a second-order linear ordinary differential equation (ODE) possessing four regular singular points, thus generalizing the hypergeometric equation, which is limited to three singular points. This broad generality makes Heun's equation a crucial tool in the study of physical systems that involve more intricate potentials, boundary conditions, or geometries, many of which naturally lead to equations of the Heun type \cite{ronveaux, slavyanov, hortacsu, olver}.

Furthermore, Heun's equation accommodates multi-pole interactions, which arise in physical scenarios where multiple distinct sources—each contributing a singularity—interact with one another. The presence of multiple singular points in the equation’s structure mirrors the contributions from these different physical sources, thus providing a natural mathematical framework for modeling such multi-pole interactions. As a result, Heun’s equation has found applications across a wide array of fields, from quantum mechanics and general relativity to condensed matter physics and string theory, where systems inherently exhibit multi-scale or multi-pole characteristics.

The general Heun equation, the canonical local form with singularities at $0, 1, a, \infty$, is written as
\br
H''(z) + (\frac{\g}{z}+ \frac{\d}{z-1}+\frac{\epsilon}{z-a}) H'(z)+ \frac{\a \b z - q}{z(z-1)(z-a)} H(z) =0, \label{heun0}
\er
where $a \in \IC\backslash\{0,1\}$ is a free parameter. It must hold 
\br
\label{condab}
\a + \b + 1 = \g + \d + \epsilon.
\er
This condition guarantees the singular point $z=\infty$ to be regular.  
In many physical applications, the accessory parameter $q \in \IC$ of the general Heun equation plays the role of a spectral parameter. When the Heun form arises from linearizing a field equation or separating variables in a spatially varying background, the admissible values of $q$
 encode quantization conditions, resonance structures, and scattering behavior through the monodromy of the solutions. The equation therefore remains under active study, not only for its intrinsic analytic complexity but also because the band-gap structure of its solutions is closely tied to the spectral theory of integrable and near-integrable nonlinear wave equations, see \cite{maier} and references therein. 

In this paper we study the spectrum of a Dirac fermion in a sine–Gordon soliton background, where the induced position-dependent mass reduces the Dirac equation to a nontrivial eigenvalue problem with spinor components obeying Heun-type equations. This framework resolves the bound-state and scattering spectra in a unified manner, making explicit their dependence on the soliton parameters and bare fermion mass. We show that the Heun formalism is essential for constructing all nonzero-energy states (valence fermions), with the full scattering data encoded in Wronskians matching local Heun solutions across the soliton. The resulting spectral characterization provides a systematic basis for analyzing soliton–fermion stability and topologically protected excitations. 

The present work builds upon Ref. \cite{arxiv} by providing a pedagogical treatment of the transformation of Fuchsian equations into the canonical Heun form. Two distinct mapping procedures are introduced, and a systematic analysis of additional parameter sets is carried out to clarify the structure and equivalence of the resulting formulations. We explain qualitatively (through additional plots not presented in  \cite{arxiv}) the incident, reflected and transmitted scattering waves, as well as the phase shifts for a set of fermion mass parameters. The physical interpretation of Dirac current conservation, and its equivalence to the topological current of the scalar field, is examined within the framework of scattering unitarity. New results regarding the Fuchsian $\rightarrow$ Heun  transformation procedure are presented and discussed in the Appendices \ref{app:transf} and \ref{app:par12}.       
 
The paper is organized as follows. In the next section, we introduce the model describing a Dirac spinor coupled to a sine-Gordon soliton. Section \ref{sec:heun} is devoted to the discussion of the Heun equation arising in the kink–spinor coupled system. In Section \ref{wronskian}, we apply the Wronskian method to implement the matching of the scattering states at the origin. This section also analyzes the fermionic bound states and the phase shifts of the spinor components after propagation through the kink background. Finally, Section \ref{sec:diss} presents the conclusions and discussions.

\section{Dirac spinor coupled to a sine-Gordon soliton}
\label{sec:diracvaccum}

When a Dirac spinor is coupled to a sine-Gordon soliton, the resulting fermion–soliton system exhibits a rich interplay between nonlinear and relativistic dynamics. The sine-Gordon soliton provides a localized, topologically nontrivial background field, while the Dirac spinor experiences this background as an effective potential with a nontrivial spatial structure. The coupled field equations will be reduced to a second-order differential equation governing the spinor components. These equations generally possess multiple singularities arising from the soliton profile and boundary conditions. Consequently, they are naturally expressible in terms of the Heun equation, the most general second-order linear ODE with four regular singular points. The emergence of the Heun structure reflects the multi-scale and multi-pole character of the fermion-soliton interaction, providing a unified analytic framework for exploring bound states, scattering, and spectral properties within this system. 

Consider the following system of equations\footnote{Our notation:  $\pa_0 = \frac{\pa}{\pa t}, \pa_1 = \frac{\pa}{\pa x}$, $\slashed{\partial} = \g^{\mu}\pa_\mu $. We use $\g_0 = \(\begin{array}{cc} 0  & i\\
-i & 0\end{array}\)$, $\g_1 = \(\begin{array}{cc} 0  & -i\\
-i & 0\end{array}\)$, $\g_5 = \g_0 \g_1 = \(\begin{array}{cc} 1  & 0\\
0 & -1\end{array}\)$,  and $\psi = \(\begin{array}{c} \psi_{R}\\
\psi_{L}\end{array}\),\,\, \bar{\psi} = \psi^{\dagger} \g_0,\,  \psi_{R} \equiv (\frac{1+\g_5}{2})\psi,\, \psi_{L} \equiv (\frac{1-\g_5}{2})\psi $.}  \cite{arxiv, prd2}
\br
\label{sla}
i \slashed{\partial} \psi = M e^{2i \b \vp \g_5} \psi,
\er
and 
\br
\label{topeq}
j^{\mu} &=&-\frac{1}{\b} \, \epsilon^{\mu\nu} \pa_{\nu} \vp,\,\,\,\,\,\,\, j^{\mu} \equiv \bar{\psi} \g^\mu \psi,\,\,\,\,  \mu= 0, 1,\,\,\nu =0,1,\,\,\,\, \epsilon^{01}=-\epsilon^{10} = 1, \epsilon^{00} =\epsilon^{11}=0.
\er
In components the Eq. (\ref{sla}) becomes 
\br
\label{xis11}
(\pa_t + \pa_x) \psi_{L} + M e^{2i \b \vp }
 \psi_{R} &=& 0\\
(\pa_t - \pa_x) \psi_{R} -  M e^{-2i \b \vp} \, \psi_{L}  &=& 0. \label{xis12}
\er
Whereas, the equation (\ref{topeq}) in components becomes
\br
\label{compo11}
j^{0}&=&\psi^\star_R \psi_R + \psi^\star_L \psi_L = -\frac{1}{\b}\, \pa_{x} \vp,\\
j^{1}&=&-\psi^\star_R \psi_R + \psi^\star_L \psi_L = -\frac{1}{\b} \, \pa_{t} \vp. \label{compo12i}
\er 
Equation (\ref{topeq}) establishes an equivalence between the Noether current $j^{\mu}$  associated with the Dirac spinor and the topological current $\epsilon^{\mu\nu} \pa_{\nu} \vp$ of the real scalar field. This equivalence is a key structural feature of certain integrable models, notably affine Toda models coupled to fermions \cite{prd2}.  
 
\subsection{Spinor zero-mode coupled to a soliton}

Let us find a static solution of the above fermion-soliton system. We assume the next Ansatz for the scalar field as   
\br
\label{exp1}
e^{2i \b \vp} =- \(\frac{\tau}{\tau^*}\)^2,\,\,\,\, \tau = 1+i e^{-2 K x} ,\,\,\,\,\,K \in \IR.
\er 
The real scalar function $\vp(x)$ is defined by taking the $``log''$ function of (\ref{exp1}) and considering its principal branch as $i \b \vp = \mbox{log}{|-(\frac{\tau}{\tau^*})^2|} + i \arg{(- (\frac{\tau}{\tau^{*}})^2)}$. The function $\vp$ can be written as
\br
\label{kink11}
\vp = \frac{2}{\b} \arctan{[\frac{1}{i}\,\frac{  \tau - \tau^{\star} }{  \tau + \tau^{\star} }]} \pm \frac{\pi}{2 \b}.
\er 
Then, one has 
\br
\label{vptau1}
\vp(x) = \frac{2}{\b} \arctan{\Big[ e^{-2 K x}\Big]} \pm \frac{\pi}{2 \b}.
\er
This is a kink(anti-kink). Note that the asymptotic values become: 1) $K<0$ one has $\vp(+\infty) = \frac{\pi}{\b} \pm \frac{\pi}{2 \b}$ and $\vp(-\infty) = \pm \frac{\pi}{2 \b}$. 2) $K>0$ one has $\vp(+\infty) = \pm \frac{\pi}{2 \b}$ and $\vp(-\infty) = \frac{\pi}{\b} \pm \frac{\pi}{2 \b}$.

These functions exhibit the topological charges
\br
\label{fractop}
Q_{k(\bar{k})} &=& \frac{\b}{2\pi} (\vp(+\infty)-\vp(-\infty)),\,\,\,\,\, k=kink \,(K<0), \,\,\,\bar{k}=antikink \,(K>0)\\
&=& \pm \frac{1}{2}  . \label{fractop1}
\er 
Therefore, the state with topological charge $Q_{kink-top}^{(I)} = + \frac{1}{2}$ is the kink and the state with $Q_{kink-top}^{(I)} = - \frac{1}{2}$ is the anti-kink. Notice that these charges are fractional, i.e. one-half of the integer $\pm 1$.  In the Fig. 1 we plot this kink type soliton for $K = -M,\,\,M=\frac{3}{2},\,\,\b = 1$ and with the minus sign in the shifting constant in (\ref{vptau1}), i.e. ($-\frac{\pi}{2 \b}$). The zero-mode spinor components become
\br
\label{0mode}
\( \begin{array}{c}
\psi_R \\
\psi_{L} \end{array}\) &=& c_1 \(\begin{array}{c}
 \frac{\zeta\, e^{-K x}}{1-i e^{-2 K x}} \\
-  \frac{e^{- K x}}{1+i e^{-2 K x}} \end{array}\),\,\,\,\,\ \zeta =  \mbox{sign} (\frac{K}{M}),\,\,\, K = \pm M.
\er
The static kink (\ref{vptau1}) and  spinor (\ref{0mode}) are solutions of the firt order system of nonlinear equations (\ref{xis11})-(\ref{xis12}) and (\ref{compo11})-(\ref{compo12i}).

Let us emphasize that the system of soliton-fermion system (\ref{sla})-(\ref{topeq}) possesses a kink type solution (\ref{vptau1}) with fractional topological charge (\ref{fractop}) coupled to the spinor bound state (\ref{0mode}). The parameter relationship $K = \pm M$ must hold in order to have this soliton-fermion  solution. One can see that the soliton's width becomes $\sim \frac{1}{2 M}$, whereas the spinor bound state charge density $j^0$ is confined inside the soliton (see Fig.1). This solution can be considered to be a zero-mode bound state, as we will explain below.

\subsection{Spinor bound states and scattering modes coupled to a soliton}
Here we consider the spinor bound states (non-zero modes ou valence fermions) and scattering states coupled to the soliton (\ref{vptau1}) written in the form (\ref{exp1}). So, let us define the parameterization
\br
\label{spbs11}
\psi   = e^{- i E_1 t}\(\begin{array}c
u(x)\\
v(x)\end{array}\),
\er
where the stationary spinor components $u(x)$ and $v(x)$ define the bound states or the scattering solutions in the presence of the soliton $\vp$, and $E_1$ is the energy of these states. 
 
So, from (\ref{xis11})-(\ref{xis12}) and (\ref{spbs11}) one can write the coupled system of static equations
\br
\label{sta11}
- E_1 u + i \pa_x u +i  M e^{-2i \b \vp} v &=&0,\\
\label{sta21}
E_1 v + i\pa_x v +i M e^{2i \b \vp} u &=&0.
\er  
From (\ref{sta21}) one can write
\br
\label{vx}
v(x) = \frac{i}{M} (\frac{1+i e^{-2 K x}}{1- i e^{-2 K x}})^2 (E_1 u - i u'(x)),
\er
where the exponential $e^{2i \b \vp}$ has been substituted using the relationship (\ref{exp1}). 

Substituting this last relationship for $v$ into (\ref{sta11}) one gets 
\br
\label{u1x}
u''(x) - 4i K \mbox{sech}(2 K x) u'(x) + (E_1^2-M^2 +4 E_1 K \mbox{sech}(2 K x)) u(x) =0.
\er

\begin{figure}
\centering
\label{fig1}
\includegraphics[width=1.5cm,scale=4, angle=0,height=4.5cm]{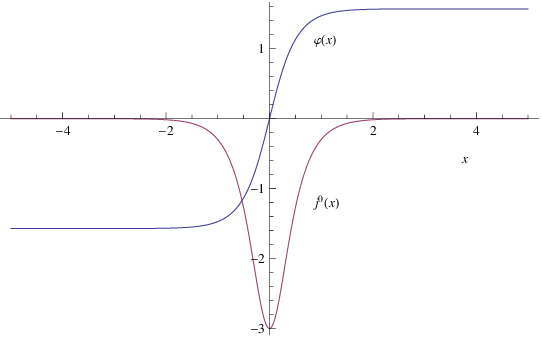}
\parbox{6in}{\caption{(color online) Scalar kink (blue) and spinor charge density (red) for $M=1.5, \b =1.$}}
\end{figure}

Note that the system (\ref{sta11})-(\ref{sta21}) can be written as an eigenvector equation 
\br
H \(\begin{array}c
u(x)\\
v(x)\end{array}\) = E_1 \(\begin{array}c
u(x)\\
v(x)\end{array}\),\,\,\,\,\, H \equiv \(\begin{array}{cc}
i\frac{d}{dx} & i M e^{-2 i \b \vp(x)}\\
-i M e^{ 2i \b \vp(x)} & -i\frac{d}{dx}\end{array}\). 
\er
Note that the Hamiltonian matrix  $H$ for real scalar field $\vp$ is a Hermitian operator, i.e.  $H^\dagger = H$, for $\b$ and $M$ real parameters. The real scalar field condition implies the spinor current density $j^\mu$ to be real in the equation (\ref{topeq}). In addition, one can show that the current is conserved. In fact, by taking the $\mu-$derivative of (\ref{topeq}) one can show 
\br
\label{currcon}
\pa_{\mu} j^\mu =0,
\er 
where the completely antisymmetric property of $\epsilon^{\mu \nu}$ has been used. Therefore from (\ref{currcon}) one can show that the charge $Q =\int_{-\infty}^{+\infty} dx\, j^0$ is a conserved quantity. This is related to the topological charge, as one can see from (\ref{topeq}) by integrating in the whole real $x-$coordinate its $0-$component
\br
\nonumber
Q &=& \int_{-\infty}^{+\infty} dx\, j^0\\
\nonumber
&=& -\frac{1}{\b} (\vp(x=+\infty)-\vp(x=-\infty))\\
&=& - \frac{2\pi}{\b^2} Q_{k(\bar{k})},\label{charges}
\er
where in the last equation it has been used the relationship (\ref{fractop}).

Some comments are in order here. 

First, the solution (\ref{0mode}) becomes the zero-mode bound state sector of the system of equations (\ref{sta11})-(\ref{sta21}), since for $E_1 =0$ these equations become the static versions of the system (\ref{xis11})-(\ref{xis12}) with the identifications $\psi_R(x,0) = u(x)$ and $\psi_L(x,0) = v(x)$ consistent with the Ansatz (\ref{spbs11}). 

Second, for the Dirac spinor interacting with the kink unitarity follows from the conservation of the Dirac current (\ref{currcon}) provided that the Hamiltonian is hermitian. 

Third, at spatial infinity the solutions reduce to free Dirac plane waves, as can be seen in the system (\ref{sta11})-(\ref{sta21}) for the scalar soliton field approaching non-vanishing constant values in the asymptotic regions $x\rightarrow \pm \infty$. In the Appendix \ref{app:par12} we discuss the asymptotic scattering and bound states. 

Fourth, unitarity implies conservation of the spatial component of the Dirac current, $j^1_{in} = j^1_{ref} + j^1_{tr}$, which amounts to the relation $R+T=1$, with reflection and transmission coefficients defined using the flux, i.e. $R = \frac{j^1_{ref}}{j^1_{in}}$ and $T = \frac{j^1_{tr}}{j^1_{in}}$. 

Fifth, in our case unitarity is tied to the topological structure of the theory due to the equation (\ref{topeq}). The fermionic probability flux is conserved as long as the topological charge associated with $\vp$ is conserved. Reflection and transmission can be interpreted as redistribution of topological flux rather than loss or gain of probability.        
  
\subsection{Heun's equation}
\label{sec:heun}
In order to motivate the transformation of the equation (\ref{u1x}) into a Heun-type equation, we can begin by analyzing the structure of this equation and identifying the key features that allow it to be recast in the form of the general Heun equation (\ref{heun0}).  To recast this equation as a Heun-type equation, we observe the following key features:

1. Singularities. The term $\mbox{sech}(2Kx)$ in (\ref{u1x}) suggests the presence of singularities at  $x = \pm \infty$; where the solution $u(x)$ will typically be localized around the soliton, and the equation will exhibit different asymptotic behavior at the boundaries of the soliton. Heun's equation, in its general form, is known for its ability to model such situations with multiple regular singular points.

2. Non-constant coefficients.
The presence of $\mbox{sech}(2Kx)$ as a coefficient in both the zero'th and first derivative terms introduces a form of ``non-uniformity'' into the equation, which is characteristic of Heun-type equations. In particular, Heun's equation is capable of describing systems where the potential or interaction term varies spatially (or temporally), as is the case here with the soliton profile.

3. Change of variables. By introducing a new variable-typically a rescaling or reparametrization such as $y=- \tanh{(2Kx)} $
 or another function that converts the $\mbox{sech}(2Kx)$ dependence into rational functions of $y$-the equation can be rewritten in a form with regular singularities located at finite points and, generally, an additional singularity at infinity. This transformation exposes the underlying Fuchsian character of the equation and places it naturally within the Heun family.

In this reformulated form, the equation inherits the characteristic analytic structure of Heun equations, including multiple singular points associated with the soliton-induced interaction scales. Recognizing the equation as Heun-type allows one to situate it within the broader analytic framework used to study systems with multi-scale potentials, nontrivial boundary behavior, and complex asymptotic structures. This connection enables the application of known Heun-function methods to analyze both local and global properties of the solutions.

4. Multi-scale structure and singularities. The localized soliton introduces distinct spatial scales, near its core and asymptotically far away, so no single asymptotic expansion can capture the solution across the entire domain. This multi-scale behavior, together with the presence of several singular points generated by the soliton profile and boundary conditions, naturally leads to a differential equation of Heun type. Mapping the problem to the general Heun equation accommodates the varying coefficients and singularity structure, allowing one to exploit established analytical tools for constructing and analyzing the fermion–soliton spectrum.

Firstly, we write this equation in a Fuchsian type second order ODE. So, as suggested above, let us make the next change of variable 
\br
\label{xy}
y = -\tanh{(2 K x)},
\er
in the  differential equation for $u$ (\ref{u1x}) in order to get
\br
u''(y) - 2 (\frac{y - i \sqrt{1-y^2}}{1-y^2}) u'(y) + \frac{E_1^2-M^2 + 4 E_1 K \sqrt{1-y^2}}{4 K^2 (1- y^2)^2} u(y)=0.
\er
Next, remove the square root with the trigonometric substitution
\br
\label{yth}
y= \cos{\theta}
\er 
in order to get
\br
u''(\theta) +(\cot{\theta} - 2 i) u'(\theta) +(\frac{E_1^2-M^2}{4K^2 (\sin{\theta})^2} + \frac{E_1}{K\sin{\theta}}) u(\theta)  =0.
\er
Furthermore, rationalize with the substitution 
\br
\label{thw}
w= e^{i \theta}.
\er
So, one gets
\br
\label{fuchsian}
u''(w) + \frac{2}{w(w^2 -1)} u'(w) +\(\frac{E_1^2-M^2}{K^2 (w^2-1)^2} - (\frac{2 i E_1}{K}) \frac{1}{w(w^2-1)}\) u(w) =0.
\er
This is a Fuchsian second order ODE with singular points at 
\br
w =0, w=1, w=-1, w= \infty.
\er
Thus, one has four regular singular points, which is the defining  feature of a Heun-type equation. 

To match the standard Heun canonical locations one performs a linear (M\"{o}bius) map in the equation (\ref{fuchsian}). We consider two cases.

{\bf The first case}. Consider the next mapping
\br
\label{wz}
z = \frac{w+1}{2}.
\er
This sends 
\br
w=-1 \rightarrow z=0,\,\,\,\, w=1 \rightarrow z=1,\,\,\,\,w=0 \rightarrow z=\frac{1}{2},\,\,\,\,w=\infty \rightarrow z= \infty.
\er
So, the third finite singularity is fixed at $a=\frac{1}{2}$. The equation (\ref{fuchsian}) becomes
\br
\label{fuchsian1}
u''(z) + [ \frac{a}{z(z-1)(z-a)}] u'(z) + [\frac{E_1^2-M^2}{4K^2 z^2 (z-1)^2 } - (\frac{i E_1}{K})\frac{1}{z(z-1)(z-a)}] u(z) =0,\,\,\,\, a=\frac{1}{2}.
\er
Next, in order to write (\ref{fuchsian1}) in the canonical form of the Heun equation (\ref{heun0}) one performs the usual gauge transformation as product of simple powers at the finite singularities. So, one has the transformation
\br
\label{tr1}
u(z) &=& f(z) \, H(z),\\
f(z)&=&z^{\mu} (z-1)^{\nu} (z-a)^{\sigma}.\label{tr11}
\er
Let us write the identity
\br
\label{id1}
u''(z) + P u'(z) + Q u(z) = f(z) [H''(z) + (P+2 \frac{f'}{f}) H'(z) + (Q+P\frac{f'}{f} + \frac{f''}{f}) H(z)], 
\er
with
\br
P &=& \frac{a}{z(z-1)(z-a)} \nonumber\\
&=& \frac{1}{z} + \frac{1}{z-1} -\frac{1/a}{z-a} \label{Pp1}
\\
Q&=& (\frac{E_1^2 - M^2}{4 K^2})\frac{1}{z^2(z-1)^2} - (\frac{i E_1}{K})[\frac{1}{z(z-1)(z-a)}] ,\,\,\,\,\, a= \frac{1}{2}. \label{Q1}
\er
and
\br
\label{f1}
\frac{f'}{f} &= & \frac{\mu}{z} + \frac{\nu}{z-1} + \frac{\s}{z-a},\\
\frac{f''}{f} &= & (\frac{f'}{f})' + (\frac{f'}{f})^2.\label{f2}
\er 
Next, taking into account the relationship (\ref{Pp1}) and the coefficient of $u'$ in (\ref{id1}), i.e. $(P+2 \frac{f'}{f})$, one can determine the parameters $\g,\d, \epsilon$ appearing in the coefficient of $u'$ of the standard Heun equation (\ref{heun0}). So, one has
\br
\label{gde1}
\g = 1+ 2 \mu,\,\,\,\, \d = 1+ 2 \nu,\,\,\,\,\,\epsilon = 2 \s -2.
\er 
Now, let us consider the coefficient of $H(z)$ in (\ref{id1}). Taking into account (\ref{Pp1})-(\ref{f2}) one can write 
\br
\label{h1}
Q+P\frac{f'}{f} + \frac{f''}{f} &=& (\frac{E_1^2 - M^2}{4 K^2})\frac{1}{z^2(z-1)^2} - (\frac{i E_1}{K})[\frac{1}{z(z-1)(z-a)}]+\\
\label{h2}
&& (\frac{1}{z} + \frac{1}{z-1} -\frac{2}{z-a} )( \frac{\mu}{z} + \frac{\nu}{z-1} + \frac{\s}{z-a}) + \\
\label{h3}
&& ( \frac{\mu}{z} + \frac{\nu}{z-1} + \frac{\s}{z-a})' + ( \frac{\mu}{z} + \frac{\nu}{z-1} + \frac{\s}{z-a})^2.
\er
Let us examine the double pole terms appearing in (\ref{h1})-(\ref{h3}) at $z=0, 1$ and $a$ and find  the relevant parameters in order to convert them into simple pole expressions. So, after some simplifications, the terms with poles at $z=0$ and $1$ can be written as
\br
\label{dp1}
(\frac{E_1^2 - M^2}{4 K^2})\frac{1}{z^2(z-1)^2} + \frac{\mu^2}{z^2} + \frac{\nu^2}{(z-1)^2} &=& \frac{(\frac{E_1^2 - M^2}{4 K^2})+ \mu^2 - 2 \mu^2 z +(\mu^2+\nu^2) z^2}{z^2(z-1)^2}\\
&=&
- \frac{k^2}{2 K^2} \frac{1}{z(z-1)}, \label{dp2}
\er
where the r.h.s in (\ref{dp2}) with simple poles is achieved provided that 
\br
\label{e12}
 E_1^2 \equiv k^2 + M^2,\,\,\,\,\, \mu =- e_1 \frac{i k}{2K},\,\,\,\, \nu = -e_2 \frac{i k}{2K},\,\,\,\,\, e_j = \pm 1, \, j=1,2.
\er
In addition, the terms with double poles at $z= a$ in (\ref{h2})-(\ref{h3}) become 
\br
-\frac{3 \s}{(z-a)^2} + \frac{\s^2}{(z-a)^2},
\er
whose total contribution vanishes provided that 
\br
\label{sig1}
\s = 0, 3.
\er   

Next, identify the $\a \b$ and the accessory parameter $q$ in the coefficient of  $H$. From (\ref{h1})-(\ref{h3}) one can write the terms involving the simple poles as 
\br
\label{sp2}
\frac{[-\frac{k^2}{2K^2} -(\mu+\nu) + 2 \s + 2 \mu \nu + 2 \mu \s + 2 \nu \s]\, z + \frac{3\mu}{2} - \frac{\nu}{2} - \mu \nu - \s - 2 \s \mu + \frac{k^2}{4 K^2}- i \frac{E_1}{K}}{z (z-1) (z-a)}.
\er
So, comparing the fraction (\ref{sp2}) with the relevant factor in (\ref{heun0}), i.e. $\frac{\a \b z - q}{z(z-1)(z-a)}$, one has
\br
\nonumber
\a \b &=&-\frac{k^2}{2K^2} -(\mu+\nu) + 2 \mu \nu + 2 \s + 2 \mu \s + 2 \nu \s,\\
\label{abq1}
&=& - (1+ e _1 e_2) k \(\frac{k - i e_1 K}{2 K^2} \) + \(-i \frac{(e_1+e_2)k}{K}  +2 \) \s\\
 q &=& -\frac{k^2}{4K^2} + i \frac{E_1}{K} - \frac{3\mu}{2} + \frac{\nu}{2} + \mu \nu + \s + 2 \mu \s,
\nonumber\\ 
&=&-\frac{(1+e_1 e_2)k^2}{4 K^2} + i e_1  \frac{(3- e_1 e_2)k}{4 K}  + \frac{i E_1}{K}+(1- i e_1\frac{k}{K}) \s.
\er
In addition, from (\ref{condab}) and (\ref{gde1}) one has
\br
\a + \b & = & 2 (\mu + \nu + \s)-1.\\
&=&  -\frac{i e_1 (1+ e_1 e_2) k}{K} + 2\s -1.\label{abq2}
\er  

Then, from the above relationships one has
\br
\a &=& -\frac{1}{2}(1+e_3- 2 \s)- i e_1  (1 + e_1 e_2)\frac{k}{2 K};\,\,\,\,\,\,\,\,\,e_a=\pm 1; \,\,\, a=1,2,3.\\
\b &=&  -\frac{1}{2}(1-e_3- 2 \s) - i e_1 (1 + e_1 e_2) \frac{k}{2 K}.
\er 
For the both values of the product $e_1 e_2 = \pm 1$ in (\ref{e12}), respectively, one can write the next sets of parameters 
\br
\label{e2par1}
e_1 e_2 =1,\,\,\,\,\,\,\,\,\,\,\,\,\,\,\,\,
\a &=& -\frac{1}{2}(1+e_3-2\s)-  e_1 \frac{i k}{K},\,\,\,\,\,\,\, \b = -\frac{1}{2}(1-e_3-2\s) -  e_1 \frac{i k}{K},  \\
\label{e2par2}
\nu &=& \mu =  -e_1 \frac{i k}{2K},\,\,\,\,
\g = \d = 1- e_1 \frac{i k}{K},\\
 q &=& -\frac{k^2}{2 K^2} + i\frac{ 2E_1+e_1 k}{2 K} +(1- i e_1\frac{k}{K}) \s;\label{e2par3}
\\
\label{e2par4}
e_ 1 e_2 =-1,\,\,\,\,\,\,\,\,\,\a &=& -\frac{1}{2}(1+e_3-2\s) ,\,\,\,\,\,\,\, \b = -\frac{1}{2}(1-e_3-2\s)  \,\,\,\,\,\,\,\,\,\,\, \mu = -\nu = - e_1 \frac{i k}{2K},\\
\label{e2par5}
\g &=& 1- e_1 \frac{i k}{K},\,\,\,\,\,\,\,\,\d = 1 + e_1 \frac{i k}{K},\\
q &=&   i \frac{E_1+ e_1 k }{ K};\label{e2par6}
\er 
In the Appendix \ref{app:case1}, Eqs. (\ref{sab1})-(\ref{sab3}) we consider the second case $e_ 1 e_2 =-1$ above, i.e. the Eqs. (\ref{e2par4})-(\ref{e2par6}).  

Therefore, the spinor components $u(z)$, which follows from (\ref{tr1})-(\ref{tr11}), can be written as 
\br
\label{uz}
u_1(z) & = &  z^{\mu} (z-1)^{\nu} Hl[a,q;\a, \b,\g,\d;z]\\
z &=& \frac{1}{1-i e^{2 K x}}, \label{uz10}
\er
where $Hl[a,q;\a, \b,\g,\d;z]$ denotes the six-parameter local Heun function \cite{ronveaux, olver, slavyanov}. The variable $z$ in (\ref{uz10}) follows after the successive variable transformations (\ref{xy}), (\ref{yth}), (\ref{thw}), and (\ref{wz}) are performed, i.e. $x \rightarrow y$, $y \rightarrow \theta$, $\theta \rightarrow w$ and $w \rightarrow z$, respectively. 

In addition, performing similar successive variable transformations, the $v_1(z)$ component can be written from the equation (\ref{vx}) as
\br
v_1(z) &=& -\frac{i}{4M} \frac{1}{(z-a)^2} \(iE_1 u_1(z) + 2K z (z-1) u_1'(z)\),\label{vz}
\er
where $u_1(z)$ is provided in (\ref{uz}).
 
For a fermionic mode incident on the kink from the left, let us consider the solution of Eq. (\ref{u1x}) in terms of the Heun's local function $Hl[a, q; \a, \b, \g, \d; z]$, such that asymptotically it must reduce to the transmitted plane wave, being  $\sim e^{ik x}$. Then, the corresponding solution becomes
\br
\label{ur10}
u_{1}^{(1)}(x) &=& z^\mu (z-1)^{\nu} Hl[a, q; \a, \b, \g,\d; z],\,\,\, \mu = - \nu =  \frac{ik}{2K}, \\
&=& e^{\frac{\pi k}{4K}} \, e^{ik x}\, Hl[\frac{1}{2}, i \frac{E_1+k}{K}; -1, 0, 1-i\frac{k}{K}, 1+i\frac{k}{K}; \frac{1}{1-i e^{2 K x}}],\label{ur1}
\er
where the notation $Hl[\frac{1}{2}, q; \a, \b, \g,\d; z]$ is used for the six parameter local Heun function \cite{ronveaux, slavyanov}.
 Note that we have used the parameter set provided in (\ref{sab1})-(\ref{sab3}) for $e_1e_2=-1, e_1=1.$ 
In fact, the behavior of $z$ in the original spatial coordinate becomes $z \rightarrow i e^{-2 K x}$ in the limit $ x \rightarrow + \infty$. So, the argument of the local Heun function 
tend to zero (one of the regular singular points at $z=0$) as $ x \rightarrow +\infty$. The local Heun function 
$Hl[\frac{1}{2}, q; \a, \b, \g,\d; z]$  is analytic at the regular singular point $z=0$ and is normalized such that 
$Hl(0)= 1$. Around this point, it admits a convergent Taylor expansion. In the complex $z-$plane the radius of convergence of this local series is given by $min\{|a|, 1\}$; for $a=1/2$, this yields a convergence radius of $1/2$.

{\bf The second case}. Consider the next transformation in the Fuchsian equation (\ref{fuchsian})
\br
\label{wz1}
z = \frac{1-w}{2}.
\er
This sends 
\br
w=-1 \rightarrow z=1,\,\,\,\, w=1 \rightarrow z=0,\,\,\,\,w=0 \rightarrow z=1/2,\,\,\,\,w=-\infty \rightarrow z= \infty.
\er
So, the third finite singularity is also fixed at $a=\frac{1}{2}$. The equation (\ref{fuchsian}) becomes
\br
\label{fuchsian11}
u_2''(z) + [ \frac{a}{z(z-1)(z-a)}] u_2'(z) + [\frac{E_1^2-M^2}{4K^2 z^2 (z-1)^2 } + (\frac{i E_1}{K})\frac{1}{z(z-1)(z-a)}] u_2(z) =0,\,\,\,\, a=\frac{1}{2}.
\er
Remarkably, this equation is similar to the first case equation in (\ref{fuchsian1}) except that the sign of the term with the factor $(\frac{i E_1}{K})$ is reversed. Therefore, in order to write (\ref{fuchsian11}) in the canonical form (\ref{heun0}) one performs similar gauge transformation as product of simple powers at the finite singularities as in (\ref{tr1})-(\ref{tr11}). Following similar steps for the case $e_ 1 e_2 =-1$ above, one can get the relevant parameters. In the Appendix \ref{app:case2}, Eqs. (\ref{km1})-(\ref{km3}) we provide the corresponding parameters.

So, in this case the spinor component $u(z)$ being a solution of (\ref{fuchsian11}), which follows from (\ref{tr1})-(\ref{tr11}), can be written as 
\br
\label{uz1}
u_2(z) & = &  z^{\mu} (z-1)^{\nu} Hl[\frac{1}{2},-i \frac{E_1+ k }{ K} ;-1, 0,1 + \frac{i k}{K},1 - \frac{i k}{K};z]\\
z &=& \frac{1}{1+i e^{-2 K x}}, \label{uz11}
\er
where  the parameters  were taken from (\ref{km1})-(\ref{km3}) with $e_1 =1$.  

In addition, performing similar successive variable transformations, the $v(z)$ component can be written from the equation (\ref{vx}) as
\br
v_2(z) &=& -\frac{i}{4M} \frac{1}{(z-a)^2} \(iE_1 u_2(z) - 2 K z (z-1) u_2'(z)\),\label{vz1}
\er
where $u_2(z)$ is provided in (\ref{uz1})-(\ref{uz11}).

Some comments are in order here. First, the behavior of the both types of solutions above are different at the asymptotic values  $x \rightarrow \pm \infty$ corresponding to the regular singularities   $z=0,1$. The argument of $u_1(z)$ tends to zero as $x \rightarrow  \infty$, whereas the argument of $u_2(z)$ tends to zero as $x \rightarrow -\infty$. 

Second, in order to get the transmission and reflection coefficients, it is needed the asymptotic behavior of $u_1$ as $x \rightarrow  -\infty$. However, the Heun function $Hl[a,q;\a, \b,\g,\d;z]$ is only analytical in the range of $|z| < min\{a, 1\}$. As  $x \rightarrow  -\infty$, we have that the argument of $u_1$ behaves as $ z=\frac{1}{1-i e^{2 K x}} \rightarrow 1$. So, $u_1$ is thus not analytical \cite{ronveaux, olver, slavyanov}. This problem
can be solved by considering another solution with good asymptotic behavior as $x \rightarrow  -\infty$ \cite{loginov, chai, chen}. The second type solution $u_2$ meets this condition, since its argument tends to zero as $x \rightarrow -\infty$, i.e. 
$z=\frac{1}{1+i e^{-2 K x}} \rightarrow 0$. As we will see below, using the both type of solutions $u_1$ and $u_2$ obtained in this way and the 
matching conditions for the fermionic wave functions, one can derive general expressions for the transmission and reflection coefficients.  

Third, since one has $\g \not\in \IZ$ ($\g \in \IC$ in general for scattering states since $\frac{k}{K} \in \IR$, see (\ref{sab3})) the Heun equation has two linearly independent solutions near $z=0$ \cite{ronveaux, olver, slavyanov}. From the relationship (\ref{tr1})-(\ref{tr11}), since $\frac{k}{K}$ is real for the scattering states, one can get two linearly independent solutions $u_1^{(1)}(z)$ and $u_{1}^{(2)}(z)$ as 
\br
\label{u1z}
u_1^{(1)}(z) & = &  z^{\mu} (z-1)^{\nu} Hl[a,q;\a, \b,\g,\d;z]\\
u_{1}^{(2)}(z) & = &  z^{\mu} (z-1)^{\nu} z^{1-\g} Hl[a,q+(\epsilon+ \d a)(1-\g);\a-\g+1, \b-\g+1,2-\g,\d;z],\label{u11z}
\er
where the parameters (\ref{sab1})-(\ref{sab3}) can be used. 

A similar construction can be performed for the second case, providing two linearly independent solutions $u_2^{(1)}(z)$ and $u_{2}^{(2)}(z)$ with parameters (\ref{km1})-(\ref{km3}). In fact, for the second type of solutions the argument of the local Heun function tend to zero as $x \rightarrow  -\infty$, meaning that the local Heun function remains analytic in this limit. Note that the behavior of $z$ in the original spatial coordinate becomes $z \rightarrow -i e^{2 K x}$ from (\ref{uz11}). Then, a first  solution of (\ref{fuchsian11}) from (\ref{uz1})-(\ref{uz11}) can be written as
\br
u_2^{(1)}(x) &=& z^\mu (z-1)^\nu\, Hl[\frac{1}{2}, -i \frac{E_1+k}{K}; -1, 0, 1+i\frac{k}{K}, 1-i\frac{k}{K}; \frac{1}{1+i e^{-2 K x}}],\label{uright1}\\
&=& e^{-\frac{\pi k}{4K}} e^{i k x} Hl[\frac{1}{2}, -i \frac{E_1+k}{K}; -1, 0, 1+i\frac{k}{K}, 1-i\frac{k}{K}; \frac{1}{1+i e^{-2 K x}}]\label{uright12}
\er
where the parameters correspond to the set (\ref{km1})-(\ref{km3}) with $e_1=1$.

The second solution becomes
\br
u_2^{(2)}(x) &=& z^\mu (z-1)^{\nu} z^{1-\g} Hl[a, q+(a \d + \epsilon)(1-\g); \a+1-\g, \b+1-\g, 2-\g, \d; \frac{1}{1+i e^{-2Kx}}], \label{ul1}\\
&=&\nonumber e^{-\frac{\pi k}{2K}} e^{- i kx} (i+ e^{2Kx})^{\frac{ik}{2K}} \times\\
&& Hl[\frac{1}{2}, -i\frac{2E_1-k}{2K}- \frac{k^2}{2K^2}; -1-\frac{ik}{K}, -\frac{ik}{K}, 1-\frac{ik}{K}, 1-\frac{ik}{K}; \frac{1}{1+i e^{-2Kx}}], \label{ul12}
\er

Fourth, in (\ref{u1z})-(\ref{u11z}) the Heun function is an inﬁnite series, $Hl[a,q;\a, \b,\g,\d;z] = \sum_{n=0}^{\infty} h_n z^n$. The coefficients $h_n$ are
determined by the three-term recurrence relationship 
\br
R_{n-1} h_{n-1} +P_n h_n +Q_{n+1} h_{n+1} = 0,
\er
such that the initial conditions 
$h_0 = 1$ and $h_{-1} = 0$ are assumed. Here $R_n = (n+\a)(n+\b), P_n = -q-n(n-1+\g)(1+a)- n(a\d+\epsilon)$, and $Q_n = a n(n-1+\g)$.

Fifth, in addition to solutions expressed in terms of Heun functions, exact closed-form solutions in elementary functions also exist under specific conditions, as examined below. It has been demonstrated that, for certain special parameter regimes
\br
\label {N1}
\a, \b &=& -N,\,\,\,\,\,N=0,1,2,...\\
h_{N+1} &=&0,
\er
the Heun function $Hl[a,q;\a, \b,\g,\d;z]$ admits a finite-series representation and truncates to a polynomial in $z$ \cite{ronveaux, slavyanov}. This truncated condition allow us to obtain exact analytical solutions. In fact, in  our case above for the solutions $u_1^{(1)}(z)$ and $u_2^{(1)}(z)$ one has $\a =-1$; so, the condition $h_2 =0$ terminates those series up to the first order in $z$.

\section{Wronskian method and matching scattering states}
\label{wronskian}

The solutions to the differential equation (\ref{u1x}) can be 
expressed in terms of the local Heun functions \cite{ronveaux, olver, slavyanov}. Note that we have used the parameter set provided in (\ref{sab1})-(\ref{sab3}) for $e_1e_2=-1, e_1=1.$ In Eq. (\ref{ur1}), the argument of the local Heun function 
tend to zero (one of the regular singular points at $z=0$) as $ x \rightarrow +\infty$. The local Heun function 
$Hl[\frac{1}{2}, q; \a, \b, \g,\d; z]$  is analytic at the regular singular point $z=0$ and is normalized such that 
$Hl(0)= 1$. Around this point, it admits a convergent Taylor expansion. In the complex $z-$plane the radius of convergence of this local series is given by $min\{|a|, 1\}$; for $a=1/2$, this yields a convergence radius of $1/2$.

Although the Taylor expansion of 
$Hl[\frac{1}{2}, q; \a, \b, \g,\d; z]$ possesses only a finite radius of convergence, the function itself admits analytic continuation to the entire complex plane, with a branch cut conventionally taken along 
$[\frac{1}{2}, \infty]$. It is thus well-defined at all finite points of the complex plane except at the regular singularities $z= 1/2$ and $z=1$. In the Eq. (\ref{ur1}), the argument of the local Heun function approaches unity as $x\rightarrow -\infty$, implying that these local solutions cease to provide a valid representation in this asymptotic regime.

However, as discussed above, for the second type of solutions the argument of the local Heun function tend to zero as $x \rightarrow  -\infty$.  Therefore, the general solution will be a linear combination of (\ref{uright12}) and (\ref{ul12})
\br
\label{u12}
u_s(x) = c_1 u_2^{(1)}(x) + c_2 u_2^{(2)}(x). 
\er 

The solutions given in Eqs. (\ref{ur1}), $u_1^{(1)}$, and $u_s$ in (\ref{u12}) possess overlapping domains of analyticity in the variable $x$
. By equating these representations $u_1^{(1)}|_{x=x_0}=u_s|_{x=x_0}$ 
together with their first derivatives $u_1^{(1)'}|_{x=x_0}=u'_s|_{x=x_0}$, one can determine the coefficients $c_1$ and $c_2$ appearing in Eq. (\ref{u12}). The specific choice of the matching point $x=x_0$ is immaterial, provided it lies within the interval $(-\infty, + \infty)$. For convenience and by symmetry, we select $x_0=0$ as the matching point. So, one has
\br
\label{match1}
u_1^{(1)}|_{(x=0)} &=& c_1 u_2^{(1)}|_{(x=0)} + c_2 u_2^{(2)}|_{(x=0)},\\
\frac{d}{dx}u_1^{(1)}|_{(x=0)} &=& c_1 \frac{d}{dx} u_2^{(1)}|_{(x=0)} + c_2 \frac{d}{dx} u_2^{(2)}|_{(x=0)}.\label{match1c}
\er
This system provides an explicit expressions
\br
\label{match11}
c_1 = \frac{W(u_1^{(1)}, u_2^{(2)})}{W(u_2^{(1)}, u_2^{(2)})}|_{(x=0)},\,\,\,\,\,\,\,\, c_2 = -\frac{W(u_1^{(1)}, u_2^{(1)})}{W(u_2^{(1)}, u_2^{(2)})}|_{(x=0)},
\er
where $W(v,w) \equiv v w'-w v'$ is the Wronskian of the two functions $v$ and $w$.

Figures 2, 3, 4, 5 and 6 display the real and imaginary parts of the scattering states corresponding to the spinor components, for $\b =1, M=5, k=2.5, E_1=+5.6$. These plots qualitatively demonstrate the effectiveness of the matching conditions imposed at $x=0$ for the relevant scattering solutions. In particular, they illustrate the role of the Wronskian method in determining the coefficients $c_1$ and $c_2$ appearing in Eq. (\ref{match11}).

The Figs. 2 show the individual components around $x=0$ of the $u-$component. Scalar kink (blue) in the left and right Figs. The real and imaginary components  of the incident, reflected and transmitted scattering wave function $u(x)$: Left Fig. $Re[c_1 u_{in}(x)]$ (green), $Re[c_2 u_{ref}(x)]$ (brown), $Re[u_{tr}(x)]$ (red). The right Fig. shows $Im[c_1 u_{in}(x)]$ (green) and $Im[c_2 u_{ref}(x)]$ (brown), $Im[u_{tr}(x)]$ (red).

In order to inspect their behavior around $x=0$ we plot the Figs. 3, 4, 5, and 6. Figure 3 depicts the scalar kink (blue) localized around the origin. The real part of the scattering wave function  $u(x)$ in the region $x < 0$ is shown as the combination
\br
Re[c_1 u_{in}(x) +c_2 u_{ref}(x)],
\er
with green line, while in the region $x > 0$ it is plotted as
\br
Re[u_{tr}(x)],
\er  
with red line. The inset highlights the smooth and continuous behavior of the 
$u$-component at the origin. This provides qualitative confirmation of the effectiveness of the Wronskian-based matching procedure imposed at 
$x=0$.

Furthermore, figure 4 depicts the same scalar kink (blue) localized around the origin. The imaginary part of the scattering wave function  $u(x)$ in the region $x < 0$ is shown as the combination
\br
Im[c_1 u_{in}(x) +c_2 u_{ref}(x)],
\er
with magenta line, while in the region $x > 0$ it is plotted as
\br
Re[u_{tr}(x)],
\er  
with red line. The inset emphasizes the smooth and continuous behavior of the imaginary component of $u$ at the origin. This qualitatively confirms the effectiveness of the Wronskian-based matching procedure applied at $x=0$.

The Figs. 5 and 6 show the real and imaginary  parts of the $v-$component. These figures show qualitatively similar behavior as the $u-$component described above.

\subsection{Fermionic bound states}

We now turn to the fermionic bound states associated with the kink background, employing both the Heun-function and Heun-polynomial approaches. An analysis of the asymptotic behavior of Eq. (\ref{u1x}) in the limit $|x| \rightarrow + \infty$ shows that the continuum (scattering) spectrum is characterized by
\br
\label{E11}
E_1^2 = M^2 + k^2, \,\,\,\, k \in \IR, 
\er
and the bound-state spectrum satisfies
\br
\label{Ebs}
E_{bs}^2 = M^2 - \kappa^2, \,\,\,\, \kappa > 0. 
\er
The origin of these dispersion relations are discussed in the Appendix \ref{app:par12}, Eqs. (\ref{disp1})-(\ref{disp2}). From a spectral perspective, bound states correspond to a purely imaginary continuation of the quasi-momentum into the complex plane, $k \rightarrow i\kappa$, and therefore appear as isolated eigenvalues lying above and below the continuum threshold at $\pm M$. Accordingly, Eqs. (\ref{vx}) and (\ref{u1x}) imply that the $u$ and $v$ components of a fermionic bound state decay as $e^{- \kappa |x|} $ as $|x|\rightarrow +\infty $.

To analyze the bound states, we examine the behavior of the scattering solutions under the analytic continuation $k \rightarrow i \kappa$. In Eq. (\ref{ur10}), the argument of the local Heun function tends to zero as $x\rightarrow + \infty$, causing the function to approach unity. Consequently, after the substitution $k \rightarrow i \kappa$ the transmitted fermionic wave exhibits the correct bound-state asymptotic behavior, decaying as $e^{- \kappa x}$ as $x\rightarrow + \infty$.

Similarly, replacing $k$ by $i \kappa$ in Eq. (\ref{ul12}) yields a reflected wave with the proper bound-state asymptotics $e^{\kappa x}$ as $x\rightarrow - \infty$.  However, the incident wave transforms into a term proportional to $e^{- \kappa x}$, which diverges in this limit. To eliminate this unphysical behavior of the incident fermionic component, the coefficient $c_1(E_1, k)$ must vanish at 
$E_1 = E_{1n}$, $k = i \kappa = i \sqrt{M^2-E_{1n}}$, where $E_{1n}$ denotes the bound state energy.

Since the coefficient  $c_1(E_1, k)$ is explicitly given by Eq. (\ref{match11}), the fermionic bound-state energy levels $E_{1n}$ are determined by the solutions of the resulting transcendental equation
\br
\label{c1110}
c_1(E_{1n}, i \sqrt{M^2 - E_{1n}}) =0,\,\,\, n=0,1,2,...
\er

\subsection{Phase shifts and Levinson's theorem}

The coefficients $c_1$ and $c_2$ encapsulate the complete information regarding the scattering of the fermionic wave by the sine-Gordon type kink. From Eqs. (\ref{ur1}) and (\ref{u12}), it follows that the asymptotic behavior of the $u-$component of the spinor wavefunction can be schematically expressed as
\br
\label{asympt1}
c_1 e^{ikx} \rightarrow  e^{ikx} + c_2 e^{\frac{\pi k}{4 K}}e^{-ikx},
\er
which describes the decomposition of the incident wave into its transmitted and reflected components.

During the scattering process, the transmitted fermionic wave component $u$ acquires a phase shift $\d_{u}$ relative to the incident wave. According to Eq. (\ref{asympt1}), this phase shift is given by
\br
\delta_u(k) = -\arg{[c_1(k)]}.
\er

\begin{figure}
\centering
\label{fig2}
\includegraphics[width=1.5cm,scale=4, angle=0,height=4.5cm]{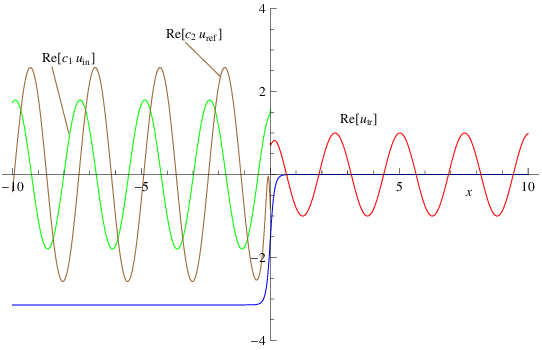}
\includegraphics[width=1.5cm,scale=4, angle=0,height=4.5cm]{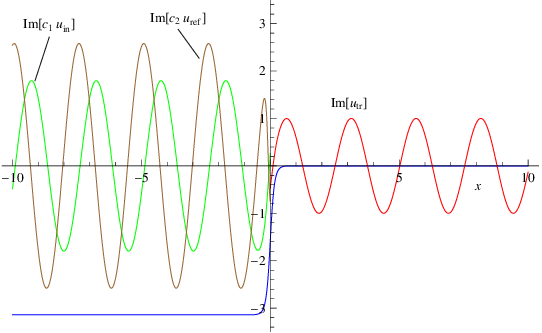}  
\parbox{6in}{\caption{(color online) Scalar kink (blue) in the left and right Figs. The real and imaginary parts of the incident, reflected and transmitted scattering wave function $u(x)$: Left Fig. $Re[c_1 u_{in}(x)]$ (green), $Re[c_2 u_{ref}(x)]$ (brown), $Re[u_{tr}(x)]$ (red), and  right Fig. $Im[c_1 u_{in}(x)]$ (green) and $Im[c_2 u_{ref}(x)]$ (brown), $Im[u_{tr}(x)]$ (red). For $\b =1, M=5, k=2.5, E_1=+5.6$.}}
\end{figure}

\begin{figure}
\centering
\label{fig3}
\includegraphics[width=1.5cm,scale=4, angle=0,height=6.5cm]{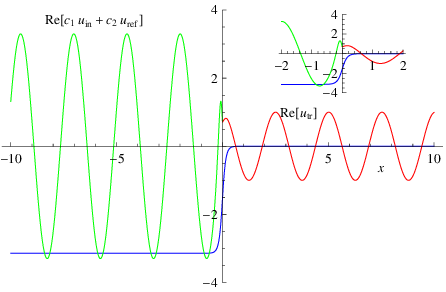}
\parbox{6in}{\caption{(color online) Scalar kink (blue) around $x=0$. The real part of the scattering wave function $u(x)$ for $x<0$, $Re[c_1 u_{in}(x) +c_2 u_{ref}(x)]$ (green), and for $x > 0$, $Re[u_{tr}(x)]$ (red). The inset figure shows the smooth matching at $x=0$.}}
\end{figure}

\begin{figure}
\centering
\label{fig4}
\includegraphics[width=1.5cm,scale=4, angle=0,height=6.5cm]{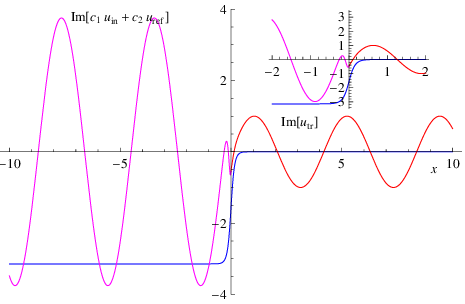}
\parbox{6in}{\caption{(color online)  Scalar kink (blue) around $x=0$. The imaginary part of the scattering wave function $u(x)$ for $x<0$, $Re[c_1 u_{in}(x) +c_2 u_{ref}(x)]$ (magenta), and for $x > 0$, $Im[u_{tr}(x)]$ (red). The inset figure shows the smooth matching at $x=0$.}}
\end{figure}

\begin{figure}
\centering
\label{fig5}
\includegraphics[width=1.5cm,scale=4, angle=0,height=6.5cm]{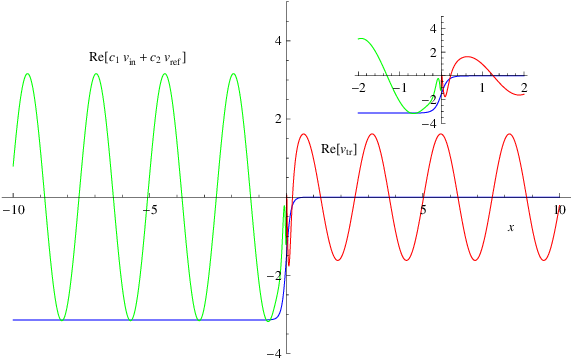}
\parbox{6in}{\caption{(color online)  Scalar kink (blue) around $x=0$. The real part of the scattering wave function $v(x)$ for $x<0$, $Re[c_1 v_{in}(x) +c_2 v_{ref}(x)]$ (green), and for $x > 0$, $Re[v_{tr}(x)]$ (red). The inset figure shows the smooth matching at $x=0$.}}
\end{figure}

\begin{figure}
\centering
\label{fig6}
\includegraphics[width=1.5cm,scale=4, angle=0,height=6.5cm]{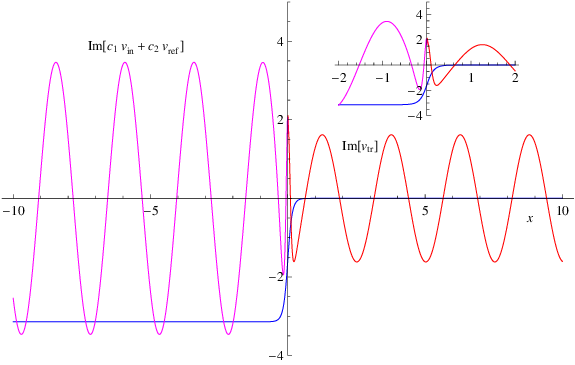}
\parbox{6in}{\caption{(color online)  Scalar kink (blue) around $x=0$. The imaginary part of the scattering wave function $v(x)$ for $x<0$, $Re[c_1 v_{in}(x) +c_2 v_{ref}(x)]$ (magenta), and for $x > 0$, $Re[v_{tr}(x)]$ (red). The inset figure shows the smooth matching at $x=0$.}}
\end{figure}

\begin{figure}
\centering
\label{fig7}
\includegraphics[width=1.5cm,scale=4, angle=0,height=5cm]{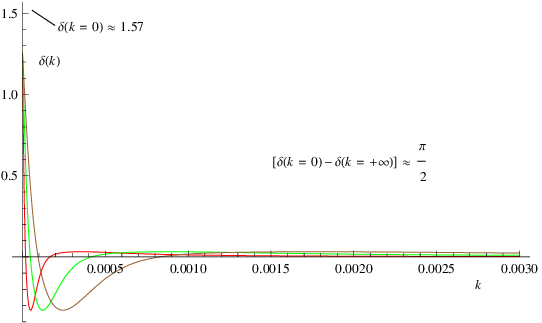}
\parbox{6in}{\caption{(color online) Phase shift $\d(k)$ vs $k$ of the scattering states $u$ and $v$ for  the fermion masses $M =2.15 \times 10^{-5}$ (red), $5 \times 10^{-5}$ (green), $10^{-4}$ (brown). Note that $\d(0)-\d(+\infty) = \frac{\pi}{2}$. }}
\end{figure}

In a similar manner, combining Eqs. (\ref{ur1}) and (\ref{u12}) and accounting for Eq. (\ref{vx}), one finds that the asymptotic behavior of the $v-$component of the spinor wave function can be schematically expressed as
\br
\label{asympt2}
c_1   e^{ikx} \rightarrow   e^{ikx} + c_2  (\frac{E_1- k}{E_1+k})\, e^{\frac{\pi k}{2 M}}e^{-ikx},
\er
such that the incident wave is decomposed into its transmitted and reflected components.

Consequently, the transmitted fermionic $v-$component acquires a phase shift $\delta_v(k)$ with respect to the incident wave. From Eq. (\ref{asympt2}), this phase shift is given by
\br
\delta_v(k) = -\arg{[c_1(k)]}.
\er
One therefore concludes that the upper and lower components of the spinor acquire identical phase shifts after scattering off the sine-Gordon–type soliton. This result contrasts with the recent findings in the ATM model with a soliton background of variable topological charge, where the phase shifts of the $u$ and $v$ components differ by a constant. In that case, a single phase shift is defined as the average of the phase shifts of the upper and lower components \cite{prd2}. In the Fig. 7 we plot the common phase shift $\delta(k)= \delta_u(k)=\delta_v(k)$ for three values of $M$. Using this Fig. one can check the Levinson's theorem \cite{barton}
\br
\label{lev1}
\d(0)-\d(+\infty) = \pi (n_b - \frac{1}{2}),
\er 
where $n_b$ is the number of bound states in a given scattering channel. One notices that the Levinson's theorem is satisfied for $n_b =1$. In the recent contribution \cite{arxiv} it has been demonstrated that the transcendental equation (\ref{c1110}) supports the bound state energies $E_{10} =0 $  and $E_{10} = \pm 0.8 M $, which are the zero and valence modes. This shows that in the positive energy channel $+ 0.8 M$ there is a bound state $n_b =1$.

\section{Discussions}
\label{sec:diss}
 
In this work, we have investigated the scattering and bound-state properties of a Dirac spinor coupled to a sine-Gordon soliton background. By formulating the spinor equations in the presence of the kink profile, we showed that the resulting differential equations for the spinor components can be expressed in terms of Heun-type equations (\ref{heun0}). This formulation provides an exact and systematic framework for analyzing the fermionic spectrum in a nontrivial topological background.

The scattering problem was addressed through the implementation of the Wronskian method, which allowed us to impose consistent matching conditions at the origin (\ref{match1})-(\ref{match1c}). This approach proved to be particularly effective in determining the scattering coefficients (\ref{match11}) and ensuring the smooth behavior of the spinor components across the kink. The numerical results for the real and imaginary parts of the scattering states qualitatively confirm the validity and robustness of the Wronskian-based matching procedure (see the Figs. 2, 3, ... 6).  

Furthermore, we analyzed the fermionic bound states by means of analytic continuation of the scattering solutions, $k \rightarrow i \kappa$. The bound-state energies were shown to be determined by the zeros of a transcendental equation arising from the vanishing of the incident-wave coefficient (\ref{c1110}). This condition guarantees the correct asymptotic behavior of the bound-state wave functions and leads to a discrete spectrum of fermionic energy levels localized around the kink.

In addition, we examined the phase shifts acquired by the spinor components during scattering off the kink background. These phase shifts encode important information about the interaction between the fermionic field and the topological soliton and may be relevant for understanding transport and polarization effects in related field-theoretical and condensed-matter systems, see \cite{arxiv} and references therein.

Heun’s equation serves as a universal mathematical structure in modern theoretical physics, encompassing phenomena from quantum to gravitational scales. It captures the essential complexity of real physical systems that lie beyond the reach of classical hypergeometric analysis and provides a powerful analytic framework for exploring nontrivial geometries, potentials, and symmetries.

The methods and results presented here can be extended in several directions. Possible future investigations include the study of other solitonic backgrounds, the incorporation of additional interaction terms, or the analysis of finite-temperature and time-dependent effects. The present framework may also find applications in effective models of low-dimensional systems where fermions interact with topological defects, as in the Ref. \cite{jhep22}.

\appendix 

\section{The sequence of transformations: $x\rightarrow y \rightarrow \theta \rightarrow w $}
\label{app:transf}
Next we provide the sequence of transformations $x\rightarrow y \rightarrow \theta \rightarrow w$ in order to get the Fuchsian second order differential equation (\ref{fuchsian}) starting from the equation (\ref{u1x}). The following identities can be written by applying successively the derivation chain-rule in each case.
    
1. The variable transformation $x\rightarrow y$: $y = - \tanh{(2 K x)}$ implies the next identities
\br
\frac{d}{dx} &=& - 2 K (1-y^2) \frac{d}{dy},\\
 \frac{d^2}{dx^2} &=& - 8 K^2 y (1-y^2) \frac{d}{dy} + 4 K^2 (1-y^2)^2  \frac{d^2}{dy^2}.
\er
2. The variable transformation $y\rightarrow \theta$: $y = \cos{\theta}$ implies 
\br
\frac{d}{dy} &=& - \frac{1}{\sin{\theta}} \frac{d}{d\theta},\\
 \frac{d^2}{dy^2} &=& - \frac{\cos{\theta}}{\sin{\theta}^3}\frac{d}{d\theta} + \frac{1}{\sin{\theta}^2}  \frac{d^2}{d\theta^2}.
\er
3. The variable transformation $\theta\rightarrow w$: $w = e^{i \theta}$ implies 
\br
\frac{d}{d\theta} &=& i w  \frac{d}{dw},\\
 \frac{d^2}{d\theta^2} &=& - \frac{d}{dw} - w^2  \frac{d^2}{dw^2}.
\er

\section{Relationships between the parameters: Cases 1 and 2}
\label{app:param}

In this appendix we discuss the relationships  between the relevant parameters. 
 
\subsection{Asymptotic scattering and bound states}
\label{app:par12}
For the scattering states one has
\br
\label{disp1}
E_1^2 = k^2 + M^2 \rightarrow E_1 = \pm \sqrt{k^2 + M^2}.
\er  
This is a dispersion relation which can be obtained from the asymptotic form of the equation (\ref{u1x}) at $x\rightarrow \pm \infty$, i.e.
\br
\label{free1}
u'' + (E_1^2-M^2) u=0.
\er
So, (\ref{disp1}) follows for the free wave solution $u \sim e^{i k x}$ of (\ref{free1}). Similarly, for the spinor bound states one has 
\br
\label{disp2}
E_{bs}^2 = -\kappa^2 + M^2 \rightarrow E_{bs} = \pm \sqrt{M^2-\kappa^2}.
\er  
It follows after substituting the expression of the form $u \sim e^{-\kappa x}\, (\kappa>0)$ for a wave with vanishing b.c. at $x\rightarrow  +\infty$ into (\ref{free1}). The threshold states occur at $\kappa = 0$ ($E_{thr} = \pm M$) and the zero modes at $\kappa = \pm M$ ($E_{bs}=0$).

\subsection{Case 1}
\label{app:case1}

The variable transformation (\ref{wz}) $w \rightarrow z$: $z = \frac{w+1}{2}$ implies 
\br
\frac{d}{dw} &=& \frac{1}{2}  \frac{d}{dz},\,\,\,\,
\frac{d^2}{dw^2} = \frac{1}{4}  \frac{d^2}{dz^2}.
\er
Using these identities one can write the Fuchsian equation (\ref{fuchsian}) in the form (\ref{fuchsian1}).  
This transformation sends $w=-1 \rightarrow z=0,\,\,\,\, w=1 \rightarrow z=1,\,\,\,\,w=0 \rightarrow z=\frac{1}{2},\,\,\,\,w=\infty \rightarrow z= \infty$. So, the regular singularities on the complex plane $z$ become at
\br
\label{sing1}
z = 0, 1, a, \infty;\,\,\,\, \mbox{with}\,\, a=\frac{1}{2}.
\er
Next, we write the parameter values which have been considered in the work.  

The canonical form of the Heun equation, given by 
(\ref{heun0})-(\ref{condab}), with the standard singular points 
in (\ref{sing1}), is specified by the parameters $\a, \b , \g , \d , q, \epsilon$. These parameters are determined by the parameter sets 
\br
\label{sab1}
\s&=&0,\,\,\, e_3= 1,\,\,\,\, e_1 = \pm 1,\\ 
\label{sab2}
\a &=& -1,\,\,\,\, \b =0, \,\,\,\,\,q =   i \frac{E_1+ e_1 k }{ K} \\
\label{sab3}
\g &=& \d^{\star} =  1- e_1 \frac{i k}{K},\,\,\,\,\mu = -\nu = - e_1 \frac{i k}{2K},
\er
where $\d^{\star}$ stands for complex conjugation of $\d$.

\subsection{Case 2}
\label{app:case2}
The variable transformation (\ref{wz1}) $w \rightarrow z$: $z = \frac{1-w}{2}$ implies
\br
\frac{d}{dw} &=& -\frac{1}{2}  \frac{d}{dz},\,\,\,\,
\frac{d^2}{dw^2} = \frac{1}{4}  \frac{d^2}{dz^2}.
\er
Using these identities one can write the Fuchsian equation (\ref{fuchsian}) in the form (\ref{fuchsian11}). 
This transformation sends 
\br
w=-1 \rightarrow z=1,\,\,\,\, w=1 \rightarrow z=0,\,\,\,\,w=0 \rightarrow z=1/2,\,\,\,\,w=-\infty \rightarrow z= \infty.
\er
So, the regular singularities on the complex plane $z$ become the same as in (\ref{sing1}). Similarly, in this case the canonical form of the Heun equation, given by (\ref{heun0})-(\ref{condab}), with the standard singular points 
in (\ref{sing1}), is specified by the parameters $\a, \b , \g , \d , q, \epsilon$. These parameters are determined by the parameter sets
\br
\label{km1}
\s&=&0,\,\,\, e_3= 1,\,\,\,\, e_1 = \pm 1. \\
\label{km2}
\a &=& -1,\,\,\,\, \b =0, \,\,\,\,\,q =  - i \frac{E_1+ e_1 k }{ K} \\
\g &=& \d^{\star} =  1 + e_1 \frac{i k}{K},\,\,\,\,\mu = -\nu = e_1 \frac{i k}{2K}. \label{km3}
\er

Finally, observe that the parameter sets (\ref{sab1})-(\ref{sab3})
 corresponding to Case 1 and (\ref{km1})-(\ref{km3}) corresponding to Case 2 are related through the transformations $E_1 \leftrightarrow - E_1$ and $e_1 \leftrightarrow - e_1$. This correspondence is consistent with the equivalence of the Fuchsian equations in Case 1, (\ref{fuchsian1}), and Case 2, (\ref{fuchsian11}), under the transformation 
$E_1 \leftrightarrow - E_1$.

\end{document}